\documentclass[sigconf]{inc/acmart}
\usepackage[]{titlesec}

\copyrightyear{2024}
\acmYear{2024}
\setcopyright{acmlicensed}\acmConference[DAC '24]{61st ACM/IEEE Design Automation Conference}{June 23--27, 2024}{San Francisco, CA, USA}
\acmBooktitle{61st ACM/IEEE Design Automation Conference (DAC '24), June 23--27, 2024, San Francisco, CA, USA}
\acmDOI{10.1145/3649329.3663515}
\acmISBN{979-8-4007-0601-1/24/06}

\acmConference[DAC '24]{the ACM/IEEE Design Automation Conference}{June 23--27, 2024}{San Francisco, CA}

\begin{document}

\title[The Magnificent Seven Challenges]{Invited: The Magnificent Seven Challenges and Opportunities in Domain-Specific Accelerator Design for Autonomous Systems}

\author{Sabrina M. Neuman}
\authornote{All authors contributed equally to this work.}
\affiliation{%
  \institution{Boston University}
  \country{}
}
\email{sneuman@bu.edu}

\author{Brian Plancher}
\authornotemark[1]
\affiliation{%
  \institution{Barnard College, Columbia University}
  \country{}
}
\email{bplancher@barnard.edu}

\author{Vijay Janapa Reddi}
\authornotemark[1]
\affiliation{%
  \institution{Harvard University}
  \country{}
}
\email{vj@eecs.harvard.edu}

\renewcommand{\shortauthors}{S. M. Neuman, B. Plancher, and V. Janapa Reddi}

\begin{abstract}
  The end of Moore's Law and Dennard Scaling has combined with advances in agile hardware design to foster a golden age of domain-specific acceleration. However, this new frontier of computing opportunities is not without pitfalls. As computer architects approach unfamiliar domains, we have seen common themes emerge in the challenges that can hinder progress in the development of useful acceleration. In this work, we present the Magnificent Seven Challenges in domain-specific accelerator design that can guide adventurous architects to contribute meaningfully to novel application domains. Although these challenges appear across domains ranging from ML to genomics, we examine them through the lens of autonomous systems as a motivating example in this work. To that end, we identify opportunities for the path forward in a successful domain-specific accelerator design from these challenges.
\end{abstract}





\maketitle

\section{Introduction}
In the first quarter of this century, computing hardware designers were faced with both the limitations of technology scaling for performance~\cite{asanovic2006landscape}, and the ensuing conflagration of on-chip power density~\cite{esmaeilzadeh2011dark}. Fortunately, the challenges of the Dark Silicon era~\cite{taylor2012dark} have transitioned into a prosperous period of specialized accelerator design~\cite{hennessy2019new,dally2020domain}.
Emerging enabling technology and tools for agile and democratized hardware design are powering an exciting \emph{land rush} in applying computer architecture to application domains that are not well-served by existing solutions.

\begin{figure}[!t]
   \centering
   \includegraphics[width=.95\columnwidth]{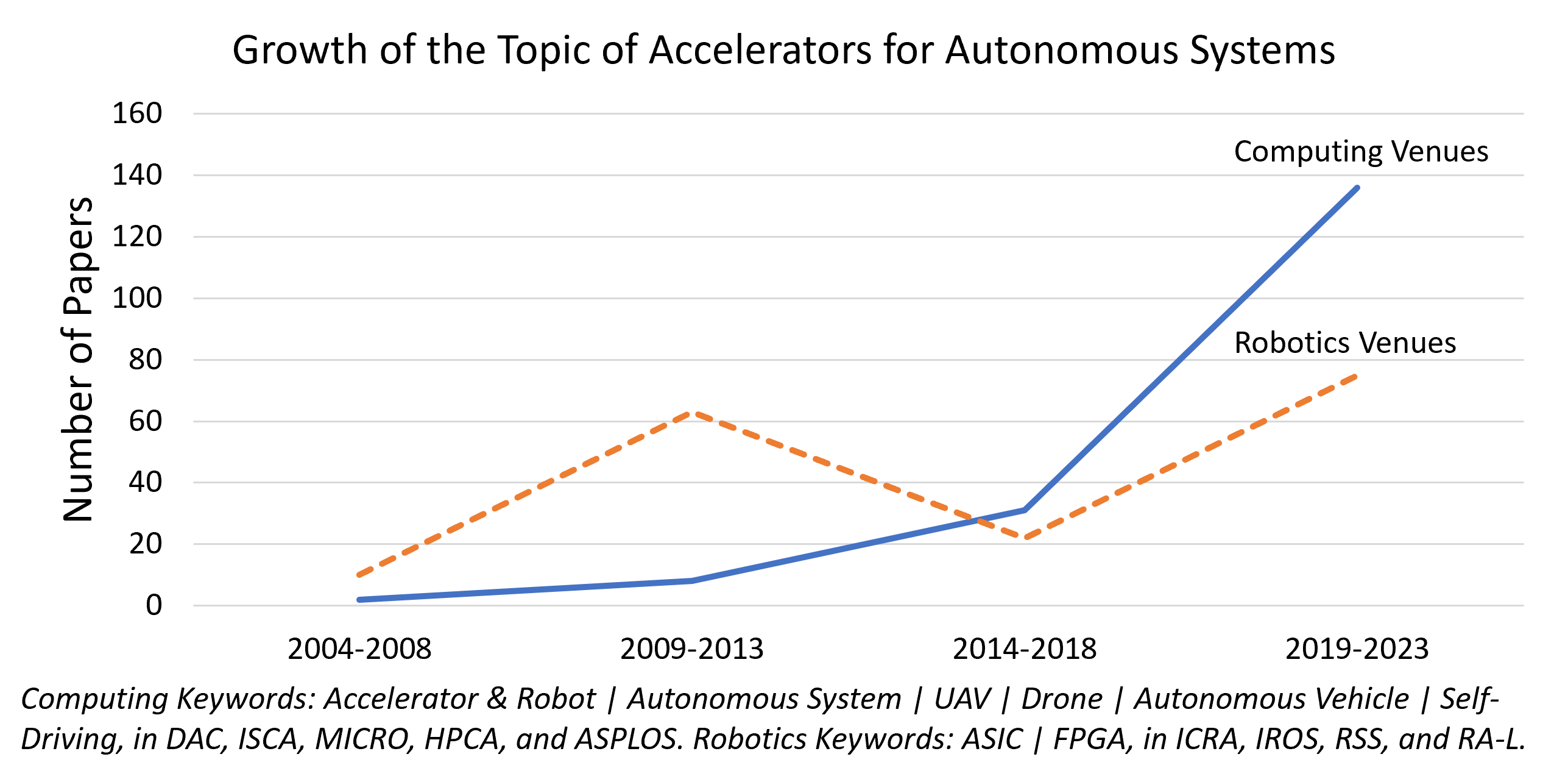} 
   \vspace{-10pt}
   \caption
   {Mention of accelerators for autonomous systems in top-tier computing and robotics venues, from Google Scholar.
   }
   \label{fig:dsa_growth}
   \vspace{-12pt}
\end{figure}

Examples of promising domains for acceleration have ranged from the popular machine learning (ML) space~\cite{esser2015backpropagation,chen2016eyeriss,chen2016diannao,parashar2017scnn,jouppi2017datacenter,ankit2019puma,gondimalla2019sparten,qin2020sigma} to emerging work in robotics~\cite{murray2016microarchitecture,sacks2018robox,suleiman2019navion,asgari2020pisces,liu2020hardware,han2020dadu,liu2021archytas,neuman2021robomorphic,bakhshalipour2022racod,neuman23RoboShape} (Fig.~\ref{fig:dsa_growth}) and genomics~\cite{turakhia2018darwin,doblas2023gmx,cali2022segram}. While these solutions have provided substantial speedups and improvements in their target applications and kernels of interest, as of yet, many of these artifacts have only had limited impact on real world production systems.
In ML, apart from the success of the Google TPU~\cite{jouppi2017datacenter}, large data centers still run the latest NVIDIA GPUs for ML training and inference.
Similarly, while early work in acceleration for robotics and genomics have delivered promising results, most commercial solutions exclusively leverage CPUs and GPUs.

In this work, we present a collection of key challenges in domain-specific accelerator design, viewed through the lens of autonomous systems, with complementary examples drawn from the more mature ML acceleration literature.
Reflecting on our experience working in this space, we find that there are common themes in the pitfalls that drive the lack of technology transfer from successful research prototypes to deployed real-world solutions.
We identify examples of current work that overcomes these challenges, providing examples of successful strategies for future work.

\section{The Magnificent Seven Challenges}
In the following, we examine each of the seven challenges in detail. We outline the pitfalls of the conventional approaches to accelerator design, and survey instructive success stories in recent work.

\subsection{Build Bridges: Engage with Domain Experts}
\label{subsec:build_bridges}
\emph{Pitfall: Interact with domains exclusively through benchmarks published in computer systems, without input from domain experts.}
\vspace{3pt}

\noindent
A common pitfall in the development of domain-specific architectures (DSAs) is the lack of direct engagement with domain experts, leading to a misunderstanding of the user's needs. This often results in the development of solutions that, while technically impressive, fail to address practical user requirements.
This mis-match can manifest \emph{across all stages of design and development}, from initially targeting the wrong bottlenecks to delivering an unusable artifact.

%
A DSA might be optimized for a specific algorithm that, unbeknownst to the developers, has little impact in the real world. 
For example, simultaneous localization and mapping (SLAM) algorithms for autonomous systems are rapidly evolving-- a 2023 survey identified $24$ representative ``active SLAM'' algorithmic approaches~\cite{placed2023survey}.
Without ongoing feedback from domain experts, an obsolete approach might be accidentally accelerated. 
In the past decade, there have been teams that have avoided this pitfall through the successful \emph{engagement of computing and robotics experts} to produce interdisciplinary work together, e.g., the Navion accelerator for visual-inertial odometry~\cite{zhang2017visual,suleiman2019navion}, and the motion planning accelerators~\cite{murray2016microarchitecture,murray2016robot} that led to the startup Realtime Robotics~\cite{realtimerobotics2024web}.
Collaborations with industry are another way to anchor design work in real-world expertise, e.g., NVIDIA's CuRobo library~\cite{sundaralingam2023curobo}. 

After design, the effective deployment of DSAs requires interfaces and wrappers optimized for existing users and workflows. 
For autonomous systems, a popular solution has been to integrate into existing open-source frameworks, e.g., 
the Robot Operating System (ROS) middleware~\cite{quigley2009ros}.
For example, a large body of work in accelerating motion planning in software (e.g.,~\cite{thomason2023motions}) has built upon the ROS-integrated Open Motion Planning Library (OMPL)~\cite{sucan2012open}.

Finally, note that designers must not only address current domain needs, but anticipate potentially \emph{unanticipated} future needs (Sec.~\ref{sec:conclusion}).
%

\subsection{Measure Twice, Cut Once: Metrics Matter}
\label{subsec:metrics_matter}
\emph{Pitfall: Only focus on improving throughput or energy-delay product.}\vspace{3pt}

\noindent
The choice of metrics for evaluating DSA performance can significantly influence the direction and outcomes of a project. 
For example, ML practitioners might prioritize accuracy, whereas accelerator designers might focus on optimizing throughput.
This discrepancy can lead to solutions that, while improving throughput, degrade model accuracy, rendering them less useful or necessitating longer training times to achieve comparable accuracy.
Beyond accuracy, pursuing traditional high performance computing metrics, e.g., tera-operations per second per watt (TOPS/W), in isolation from \emph{system-level metrics} (e.g., off-chip bandwidth) has been demonstrated to be misleading in ML accelerators~\cite{sze2020evaluate}.
Autonomous systems have analogous complex tradeoffs, for example, depending on control algorithm and hyperparameter choices~\cite{moll2021hyperplan}.
Metrics measuring, e.g., time-to-accuracy and holistic system performance, rather than raw throughput or energy efficiency, are needed. 
%

\subsection{``Widgetism'': Avoid Over-Specialization}
\label{subsec:wigitism}
\emph{Pitfall: A cycle of pick one slow algorithm, lower it to an ASIC, repeat.}\vspace{3pt}

\noindent DSAs have the potential to revolutionize computing by providing highly optimized solutions for specific domains. However, there is a significant risk of falling into the trap of ``widgetism,'' where the focus becomes too narrow and specialized, limiting the broader impact and adaptability of designs. 
The key to overcoming this
is to identify and support \emph{cross-cutting kernels} that represent fundamental operations that are applicabile across tasks (e.g., motion planning in autonomous systems~\cite{sucan2012open}).
%
In ML, for example, hardware acceleration of common kernels, such as sparse tensor algebra~\cite{dave2021hardware}, can support a wide variety of different user tasks with the same accelerator architectures.
By focusing on such 
kernels, DSAs can provide significant performance improvements while maintaining flexibility.
Over-specialization also occurs when DSAs are only \emph{evaluated} on a narrow set of tasks. 
This incentivizes high-performance ``widgets'' that may be overfit to unrealistic or uncommon use cases. 

\subsection{Pump the Brakes: Do Not Always Accelerate}
\label{subsec:pump_the_breaks}
\emph{Pitfall: Assume accelerators always improve total system performance.}\vspace{3pt}

\noindent
The drive to accelerate every aspect of computation without considering the broader system context can lead to inefficiencies, particularly in complex environments. 
Accelerators, while powerful, are not free: they consume shared resources and can introduce complexities in system scheduling and resource allocation~\cite{sze2020evaluate}. 
It is crucial to evaluate the impact of accelerators within the context of the entire system. 
For example, recent work on choosing onboard processors for UAVs~\cite{krishnan2022automatic} found that for overall mission performance, it was important to balance the processing rates of compute with sensor input rates-- and, in fact, \emph{over-provisioning compute could have disastrous effects} on the weight and battery life of the total system.
For complex autonomous systems, in some cases, the best decision might be to throttle, or even forego acceleration in favor of maintaining system harmony and efficiency.

\subsection{Chips and Salsa: Acceleration Beyond ASICs}
\label{subsec:chips_and_salsa}
\emph{Pitfall: Focus on ASICs, leaving software, GPUs, and FPGAs behind.}\vspace{3pt}

\noindent
Software solutions on CPUs and GPUs, and firmware on FPGAs are all in deployment in real commercial systems---we must not overlook opportunities to accelerate autonomous systems with these platforms.
Programmmable devices like GPUs and FPGAs offer flexibility and have benefited a wide range of applications.
Notably, utilization of GPUs has enabled rapid growth in AI and ML in the past decade.
In autonomous systems, recent work on software-only optimizations that leveraged vectorization on the CPU achieved up to 500x speedups over state-of-the-art for certain motion planning problems~\cite{thomason2023motions}.
Work on software libraries for autonomous systems algorithms (e.g.,~\cite{plancher2022grid}) can extend similar benefits to GPUs.
Even as new domain-specific ASICs are introduced into existing systems, they will likely be part of a complete heterogenous computing solution, balancing the performance of dedicated hardware and the flexibility afforded by software and programmable platforms.

\subsection{Forest vs. Trees: Take an End-to-End View}
\label{subsec:forest_vs_trees}
\emph{Pitfall: A narrow scope: acceleration begins and ends with compute.}\vspace{3pt}

\noindent
Focusing exclusively on optimizing individual kernels (even cross-cutting ones) without considering the end-to-end system \emph{beyond the computing system} (e.g., including data, I/O, sensors, actuators, and real-world effects like reliability and robustness to noise) can lead to DSAs that improve theoretical performance but fail to deliver real-world benefits.
This effect has been seen in the ``AI Tax'' of datacenter-scale ML systems, where costs of user data marshalling and network I/O transfer are substantial compared to edge device inference.
Autonomous systems are also complex to deploy, and considerations such as weight, battery technology, sensor and actuator quality, are layered on top of traditional computing design concerns~\cite{krishnan2022automatic,hadidi2021quantifying,neuman2022tiny}.
An example of a promising approach to tackling this full deployment complexity is to build simulation and testing environments that model the full system and its interactions with the environment, e.g., the MAVBench and RoSE simulators for UAV acceleration~\cite{boroujerdian2018mavbench,nikiforov2023rose}, and ILLIXR for AR/VR platforms~\cite{huzaifa2021illixr}.

\subsection{Design Global: Sustainability and Impact}
\label{subsec:design_global}
\emph{Pitfall: Design compute in isolation from its global and societal impact.}\vspace{3pt}

\noindent
Accelerator design has often viewed computing devices as isolated entities, focusing on performance metrics within the confines of the hardware itself.
However, we must consider the environmental and societal impact of computing design at scale.
For example, recent work
makes a compelling case that autonomous vehicles should be viewed as ``datacenters on wheels''~\cite{sudhakar2022data}
and predicts that future global-scale self-driving systems may require computational resources that exceed traditional datacenters. 
Similarly, recent ML research noted that the higher efficiency of cloud-based training compared to on-device processing implies that choosing to train ML models on edge devices can lead to a greater increase in carbon emissions~\cite{patterson2024energy}.
These results indicate that the design of accelerators deployed to the edge must take into consideration global factors in their manufacturing, deployment scale, and operation. 

\section{Future Directions \& Opportunities}
To address these key challenges, domain-specific accelerator design requires close collaboration between domain experts and hardware designers, adaptability to evolving algorithmic landscapes, and a holistic view of the entire system, including global impact at scale.

\subsection{Enabling Technologies and Methodologies}
\label{subsec:enabling_tools}

\noindent\textbf{Agile Design Tools.}
One key opportunity lies in the development of agile design tools that streamline the creation and customization of DSAs.
There has been a growing body of work on hardware compiler frameworks and high-level synthesis (HLS)~\cite{eldridge2021mlir,nigam2021compiler,winterstein2013high,nikhil2004bluespec}, 
and these tools should continue to be leveraged and augmented to power domain-specific design.
To be effective in the autonomous systems space,
these tools should operate at a high level of abstraction, allowing users to specify accelerator functionality through intuitive methods such as domain-specific languages (DSLs) or graphical interfaces, enabling domain experts to directly participate in the design process. 
Moreover, agile design tools should strive to bridge the gap between high-level algorithmic specifications and low-level hardware implementations, e.g., with formal verification techniques to prove correctness of generated hardware from a specification-- which is crucial in the safety-critical field of autonomous systems.
%

\noindent\textbf{End-to-End Modeling and Simulation.}
To design effective DSAs, it is essential to consider the entire system in which they will be deployed, rather than focusing solely on the accelerator in isolation.
This requires end-to-end modeling and simulation tools that can capture the complex interactions between the accelerator, the rest of the computing system, and the physical environment.
A key aspect of end-to-end modeling is the ability to accurately simulate the data flow through the system, from sensors and input devices to the accelerator and back to actuators or output devices (e.g., the MAVBench~\cite{boroujerdian2018mavbench} and RoSE~\cite{nikiforov2023rose} simulators).
This simulation should take into account the bandwidth and latency constraints, processing delays, and power and thermal characteristics of the system.
%
Finally, end-to-end modeling should consider the physical environment in which the system will be deployed. For example, for an autonomous vehicle, the model should take into account factors such as vehicle dynamics, sensor noise, and weather conditions.

\noindent\textbf{Machine Learning for System Design.}
ML
has the potential to support the design and optimization of accelerators for autonomous systems and other domains.
By leveraging the ability of ML models to learn complex patterns and make data-driven decisions, it is possible to automate many tedious and time-consuming aspects of the accelerator design process.
One promising use of ML is design space exploration. Given an algorithm specification and a set of hardware constraints, an ML model can be trained to search the space of possible hardware configurations and identify the most promising candidates considering the full-system~\cite{krishnan2022automatic}, saving time and discovering unconventional designs that experts might overlook. 

\subsection{Fostering a Robust Research Ecosystem}


\noindent\textbf{Cross-Domain Collaboration.}
Accelerator design is inherently  multidisciplinary.
To facilitate collaboration between diverse communities, it is essential to establish dedicated forums for knowledge sharing and exchange including specialized interdisciplinary workshops, conferences, research centers, and consortia.
Domain experts can also be more easily integrated into the design process through the inclusive design of tools and interfaces such as DSLs (Sec.~\ref{subsec:enabling_tools}).


\noindent\textbf{Open-Source Resources.}
A key element of a robust research ecosystem is the availability of open-source resources, including software tools, hardware designs, and datasets. 
Such resources lower the barriers to entry for new researchers, enable reproducibility and comparability of results, accelerating innovation. 
Open-source datasets and benchmarks 
that are representative of real-world workloads 
enable researchers to evaluate their designs under realistic conditions and compare their results against those of other researchers.
Examples such as the ILLIXR AR/VR workload suite~\cite{huzaifa2021illixr} and the RoSE framework for RTL and physics simulation~\cite{nikiforov2023rose} demonstrate the open-sourcing of software to fairly evaluate the performance and power efficiency of accelerator designs.


\noindent\textbf{Standardized Benchmarks and Metrics.}
To enable meaningful comparison and evaluation of accelerator designs, it is essential to establish widely-accepted, standardized benchmarks and metrics. 
These benchmarks and metrics should not only evaluate domain performance, but also energy efficiency, cost, and other key characteristics of accelerator designs.
One example of such a benchmark suite for ML accelerators is MLPerf~\cite{reddi2020mlperf}, which provides a set of reference workloads and evaluation criteria for measuring the performance of ML training and inference systems.
Finally, standardized benchmarks and metrics can also play a key role in driving innovation and guiding research priorities by identifying the most important and challenging workloads and use cases.

\subsection{Sustainable \& Responsible Hardware Design}

\noindent\textbf{Strategic Deployment of Accelerators.}
While accelerators can provide significant benefits for specific workloads, it is important to  consider their deployment scenarios (Sec.~\ref{subsec:design_global}).
This includes factors such as the cost of development and deployment (e.g.,~using emissions modeling tools~\cite{gupta2023architectural}), the scalability of the solution, and the potential for reuse across applications and domains.
This helps avoid the proliferation of short-lifespan, over-specialized hardware, which contributes to electronic waste.
For example, chiplet technology can help by providing a modular design approach that promotes the reuse and repurposing of hardware components~\cite{sudarshan2023greenfpga}. 

\noindent\textbf{Lifecycle Analysis \& End-of-Life Management.}
Lifecycle analysis (LCA) is useful for assessing the environmental impacts of a product or system throughout its entire lifespan, from raw materials extraction to end-of-life management.
By conducting LCAs for accelerator designs, we can identify opportunities for reducing the environmental footprint of hardware, such as selecting low-impact materials, optimizing manufacturing processes, or designing for easier disassembly and recycling.
Effective end-of-life management is also key to minimize the environmental impact of accelerators, from recycling, to recovering valuable materials, to the safe disposal of hazardous materials.
Expanding the scope of LCA to encompass entire autonomous systems (e.g., self-driving vehicle fleets~\cite{sudhakar2022data}) would be a step towards sustainable automation in the future.
\section{Conclusion}
\label{sec:conclusion}

By understanding the common challenges of domain-specific design for autonomous systems and beyond,
we can deploy accelerators that provide a positive broader impact both on their target applications as well as the larger world.
Finally, it is important to remember that Amdahl's Law is a moving target.
Recalling Henry Ford's possibly apocryphal~\cite{vlaskovits2011henry} remark on innovation\footnote{The automobile tycoon is often credited with the quote, “If I had asked people what they wanted, they would have said faster horses.”}, accelerator designers should recognize that \emph{anticipating the future needs} of a domain requires a constant re-examination of the fundamental benchmarks and assumptions, and dynamic analysis to continually identify new opportunities over time.
Incorporating feedback mechanisms into the design process ensures that useful contributions continue to be made as societal needs and technological landscapes evolve.

\bibliographystyle{inc/IEEE_ref}

\bibliography{inc/base,refs.bib}

\end{document}